\documentclass[referee]{raa}
\usepackage{graphicx,times}
\usepackage{natbib}
\usepackage{amssymb,amsmath}
\bibpunct{(}{)}{;}{a}{}{,}

\begin{document}

\title{Radial velocity measurements from LAMOST medium-resolution spectroscopic observations:}
\subtitle{A pointing towards the {\sl Kepler} field}

 \volnopage{ {20XX} Vol.\ {X} No. {XX}, 000--000}
   \setcounter{page}{1}

   \author{Nian Liu\inst{1}, Jian-Ning Fu\inst{1,}\footnote{Corresponding authors}, Weikai Zong\inst{1,}$^{\star\star}$,
   Jianrong Shi\inst{2,3},  Ali Luo\inst{2,3}, Haotong Zhang\inst{2}, Xiangqun Cui\inst{4}, Yonghui Hou\inst{3,4},
      Yang Pan\inst{1}, Xinrui Shan\inst{1}, Jianjun Chen\inst{2}, Zhongrui Bai\inst{2,3}, Jianxing Chen\inst{1}, Bing Du\inst{2}, Wen Hou\inst{2}, Yuchen Liu\inst{1},
      Hao Tian\inst{2}, Jiangtao Wang\inst{1}, Jiaxin Wang\inst{1}, Kefei Wu\inst{2}, Yuzhong Wu\inst{2},  Hongliang Yan\inst{2,3}
      and Fang Zuo\inst{2,3}}

   \institute{Department of Astronomy, Beijing Normal University, Beijing~100875, P.~R.~China;
   {\it jnfu@bnu.edu.cn,weikai.zong@bnu.edu.cn}\\
        \and
          Key Lab for Optical Astronomy, National Astronomical Observatories, Chinese Academy of Sciences, Beijing 100012, P.~R.~China\\
           \and
         School of Astronomy and Space Science, University of Chinese Academy of Sciences, Beijing, 100049, China\\
          \and
Nanjing Institute of Astronomical Optics \& Technology, National Astronomical Observatories, Chinese Academy of Sciences, Nanjing~210042, P.~R.~China \\
\vs \no
   {\small Received 2018 Month Day; accepted 20XX Month Day}
}

\abstract{ Radial velocity is one of key measurements in understanding the fundamental properties of stars, stellar clusters and the Galaxy.
A plate of stars in the {\sl Kepler} field were observed in May of 2018 with the medium-resolution spectrographs {of LAMOST}, aiming to test the performance of this new system which is the upgraded equipment of LAMOST after the first five-year regular survey. We present our analysis on the radial velocity measurements ({\sl RVs}) derived from these data. The results show that {slight and significant} systematic errors exist among the {\sl RVs} obtained from the spectra collected by different spectrographs and exposures{, respectively}. After correcting the systematic errors with different techniques, the precision of {\sl RVs} reach{es $\sim1.3$, $\sim1.0$, $\sim0.5$ and $\sim0.3$\,km/s} at $S/N_r=10$, 20, 50, and 100, respectively. Comparing with the {\sl RVs} {of the} standard stars {of the} APOGEE {survey}, our {\sl RVs} are calibrated with a zero-point shift of {$\sim7$\,km/s}. The results indicate that the LAMOST medium-resolution spectroscopic system may provide {\sl RVs} in a reasonable accuracy and precision for the selected targets.
}

   \authorrunning{Liu, Fu, and Zong, et al. }            
   \titlerunning{MRSLK-project spectra}  
   \maketitle

%
\section{Introduction}           
\label{sect:intro}

{The measurements of radial velocities} ({\sl RVs}) of a {large} number of stars play{s} an important role in understanding the structure of the Galaxy \citep[e.g.,][]{1998gaas.book.....B} and the kinematics of  globular clusters \citep[e.g.,][]{1979AJ.....84..752G}. {\sl RV}s are also valuable for the discovery and determination of orbital parameters of binary systems \citep[e.g.,][]{2002ApJS..141..503N}. In recent years, many large surveys provide {\sl RV}s for large samples of stars with high-precision, such as  the Sloan Digital Sky Survey for millions of stars \citep{2015ApJS..219...12A,2011AJ....142...72E,2006ApJS..162...38A} and the {\sl Gaia} observations on some seven {millions of} sources with median {\sl RV}s \citep{2018A&A...616A...1G}.

{When combined} with photometric observations, {\sl RV} variations can offer more precise constraints on the theoretical frameworks of stellar pulsation models \citep{2013ApJ...768L...6M} and present an unbiased mass determination of the components of eclipsing binary stars \citep[e.g,][]{2007A&A...471..605V}. The {\sl Kepler} {space mission} monitored about 200\,000 stars in the {region of the} constellations Cygnus and Lyrae for a period of $\sim 4$\,yr continuously \citep{2010Sci...327..977B}, providing unprecedented high-quality photometric data for many types of variable stars \citep{gilliand10,2011AJ....141...83P,2016A&A...585A..22Z}. Consequently, to fully exploit the science as offered from these {photometric observations}, different groups have been organized to provide ground-based spectra as follow-up programs, for instance, APOKASC \citep{2014ApJS..215...19P,2018ApJS..239...32P} and {the} LAMOST-{\sl Kepler} (LK) project \citep{2015ApJS..220...19D,zong2018}, { providing} {\sl RV}s for thousands of stars. Nevertheless, multiple {visits} of specific targets show particular interests in exoplanets or binary detection from periodic {\sl RV} variations \citep[see, e.g., MARVELS in][]{2008ASPC..398..449G}. The LK-project also provides multiple ($>4\times$) {\sl RV}s for about 500 stars \citep{zong2018}.

LAMOST\footnote{The Large Sky Area Multi-Object Fiber Spectroscopic Telescope (also called Gou Shoujing Telescope) which is located at the Xinglong Observatory, P.~R.~China. More details can be found in \citet{2012RAA....12.1197C} and \citet{2012RAA....12..723Z}.} is an ideal instrument for spectroscopic {observation surveys}, which can monitor more than three thousands targets per exposure \citep{1996ApOpt..35.5155W,1998SPIE.3352..839X}, vastly reducing time consumption {to measure} {\sl RV}s {for a large} number of targets.
From the pilot and the first 5-yr regular survey, LAMOST obtained more than nine million low-resolution ($R\sim1800$) spectra \citep[see, e.g.,][]{2015RAA....15.1095L}. Since 2017 September, {LAMOST was} tested with medium-resolution ($R\sim7500$) {spectrographs} with two arms covering the wavelength ranges of 630--680\,nm and 495--535\,nm, respectively \citep{zong2018}. The bright moon nights in each lunar month are reserved to {perform these test observations}.

In this {paper}, we will address an estimation of the precision of {\sl RV}s derived from the current LAMOST pipelines. It is evaluated through time-series spectroscopic observations pointing {towards} the {\sl Kepler} field. The structure of this {paper} is organized as follows. {T}he details of observations and data reduction {are} described in Section~2. We present the techniques to estimate the precision of {\sl RV}s in Section~3, {followed by the} the comparison with APOGEE {\sl RV}s in Section~4. We give our discussion in Section~5 and conclude our results in Section~6.

\section{Observations and data reduction}
\label{sect:Obs}
\subsection{Observations}

\begin{table}
\bc
\caption[]{Detailed contents of the LK07 footprint which had been observed by LAMOST equipped with medium-resolution spectrograph during 2018 May.\label{t1}}
\setlength{\tabcolsep}{5pt}
 \begin{tabular}{cccccr}
  \hline\noalign{\smallskip}
Observation date & Begin &  End &  Exposure time &  Seeing   &  Parameter \\
                          &   (UT)      &   (UT)    &                        & (arcsec)  & \\
    \hline\noalign{\smallskip}
2018 May 24  &  18:26:16   & 19:55:33     &  900\,s $\times$ 5     & $\sim$3.0   & 7214  \\
2018 May 28  &  17:23:20   & 19:39:33     &  900\,s $\times$ 7     & $\sim$2.6   & 10375  \\
2018 May 29  &  17:36:44   & 19:38:12     &  600\,s $\times$ 9     & $\sim$2.3   & 12329 \\
2018 May 30  &  17:58:56   & 19:29:23     &  900\,s $\times$ 5     & $\sim$2.4   & 7414  \\
2018 May 31  &  18:02:13   & 19:32:49     &  1200\,s $\times$ 4    & $\sim$2.3   & 6088  \\
  \noalign{\smallskip}\hline
Total              &                   &                       & 25500 s                    &    & 43420 \\
  \noalign{\smallskip}\hline
\end{tabular}
\ec
\tablecomments{1\textwidth}{The time between begin and end includes the readout time but not the overhead time.}
\end{table}

\begin{figure}
\centering
\includegraphics[width=8cm]{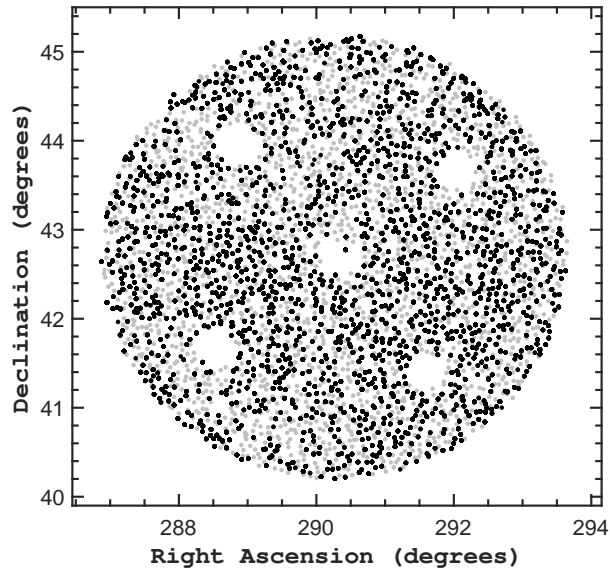}
\caption{Sky coverage of all targets (in grey) observed by LAMOST pointing towards the LK07 field. The stars atmospheric parameters derived from LASP are marked in dark.
\label{f1}}
\end{figure}

LAMOST has a focal plane of 5 degrees in diameter, equipped with 4000 fibers, {hence} the telescope can observe 4000 targets (including sky light) per exposure. One circular field in {\sl Kepler} field, LK07, had been chosen to be observed, with an aim to test the precision of {\sl RV}s from the medium-resolution spectra. More details of the classification of each
{\sl Kepler} field can be {found} in \citet{2015ApJS..220...19D}. The central position of LK07 is defined by the coordinates of {the} bright star HIP\,95119 with $V=7.03$, $\alpha (2000) =19:31:02.82$, and $\delta(2000) = +42:41:13.06$. This star is used for calculation of wavefronts to reshape the {mirrors} into good condition. The input targets are chosen based on several criteria as follows, with priority decreasing: two pulsating stars {showing} particular interests, 6 standard stars, 164 eclipsing {binary} stars, the rest stars with {a} similar strategy {as} in \citet{zong2018}. Figure\,\ref{f1} shows the spatial distribution of 3626 targets which are finally allocated {to} fibers.

Table\,\ref{t1} lists the details of the observations of that plate. {The} footprint had been observed by LAMOST from 2018 May 24, 28 to 31, {on five individual} nights. This field is given a very high priority to be observed since the {\sl Kepler} field can only be reached during the summer season \citep[see details in][]{2015ApJS..220...19D,zong2018}. {Observations can almost start} when the LK07 field enters the view of LAMOST, which is confined within two hours before and after the meridian of the central star. The overhead time to prepare for exposure is typically 30 minutes, depending somewhat on the telescope performance and weather conditions. The readout time is about 4 minutes for each exposure. When the exposure is ready, the footprint will be observed continuously until it leaves the view of LAMOST or the twilight is too bright to continue the observation. The latter one is the main reason for stopping the {observations} in late May. During the observations, the weather {condition} is typically with a seeing of around 2.5 arcsecs. A total of 30 plates have been obtained with exposures of 900\,s\,$\times$5, 900\,s\,$\times$7, 600\,s\,$\times$9, 900\,s\,$\times$5 and 1200\,s\,$\times$4. The total exposure time corresponds to 7.08 hours.

\subsection{Data reduction}
The raw products {of LAMOST observations are} the two dimensional (2D) CCD frames. For each exposure, a total of 32 (16 blue and 16 red) 2D frames are obtained, with each frame containing 250 raw spectra {almost equally spread on the CCD}. The first procedure to reduce those raw data is to evaluate the quality of observations and the telescope performance, such as seeing, cloud coverage and checking of polluted light. The 2D frames with good quality {are} used to produce 1D calibrated spectra by the LAMOST 2D pipeline, which is implemented with procedures similar to those of SDSS \citep{2002AJ....123..485S}. The main tasks of the LAMOST 2D pipeline include dark and bias subtraction, flat field correction, spectral extraction, sky subtraction, and wavelength calibration \citep[see more details in][]{2015RAA....15.1095L}. {One notes} that the 2D pipeline conducted on the medium-resolution does not contain stacking {of} sub-exposures and combining of different wavelength bands {with these procedures were used for the} low-resolution spectra.

{The scientific quality of the obtained 1D spectra is evaluated before the atmospheric parameters are calculated}. We use the signal-to-noise in {\sl SDSS}-like {\sl r} band (hereafter $S/N$ for simplification) as the indicator. The spectra with $S/N$ higher than 10 will be fed to the 1D pipeline to derive the LASP parameters and to classify the spectral type.
The {\sl RV}s for stars and redshifts for galaxies (or quasi-stellar objects) are also provided through this pipeline. The current version v2.9.7 pipeline is used for the medium-resolution spectra obtained from the LK07 plates. More details of these pipelines can {be found in} \citet{2012RAA....12.1243L} and \citet{2015RAA....15.1095L}.

\section{Analysis of Radial velocities}
\subsection{Distributions of {\sl RVs} measurements} \label{s:drv}
\begin{figure}
\centering
\includegraphics[width=10cm]{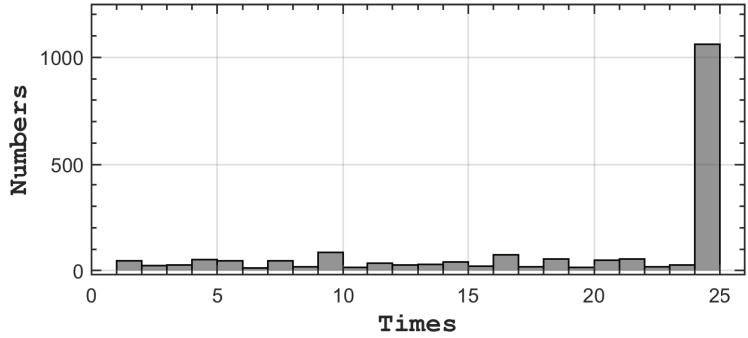}
\caption{Distribution of the times for stars derived with {\sl RV}s from the 25 exposures.
\label{expo}}
\end{figure}

\begin{figure}
\centering
\includegraphics[width=10cm]{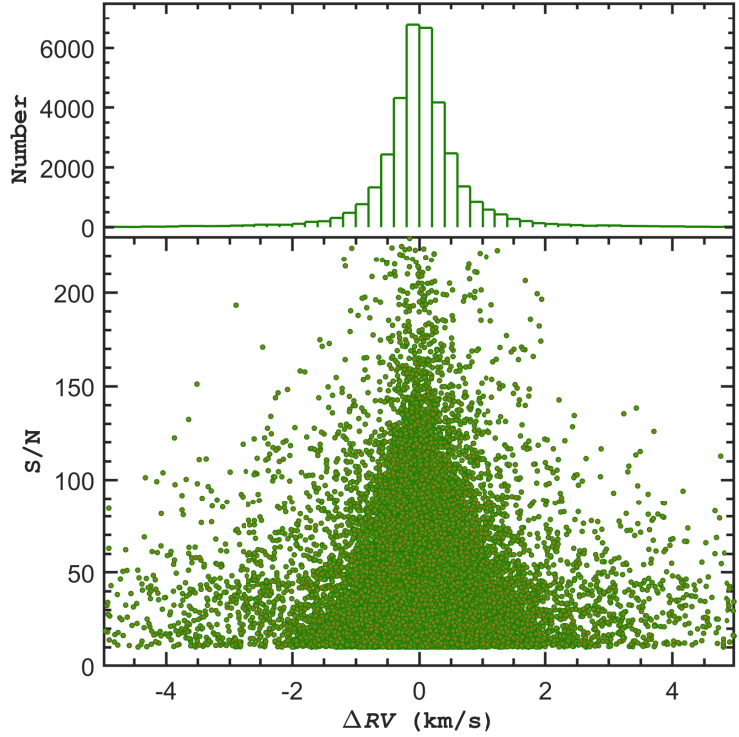}
\caption{Distribution of the relative {\sl RV}s ($\Delta RV$)
as a function of the spectra quality $S/N$ ({\sl bottom panel}).
The projection of the $\Delta RV$ histogram with a bin width of 0.2 km/s {is shown in the} {\sl top panel}. We note that the long side wings are not shown in this plot.
\label{4W8}}
\end{figure}

The calibrated spectra with high quality can definitely produce atmospheric parameters. However, we will merely discuss the results of {the} measurement of {the} precision of {\sl RV}s in this {paper}.
The {total number of {\sl RV} measurements} {obtained from the 30 plates} is 43420. The
last column of Table\,\ref{t1} lists the {individual} number of {\sl RV}s in each {night}.
We {measured typically} around 1500 {\sl RV}s from each {plate}. {We note that a Scandium arc\footnote{The Scandium will not be used any longer as a result of comparison to the Thorium-Argon arc.} was used to calibrate the wavelength for the spectra of the first 5 exposures, while, a Thorium-Argon arc was used for the rest { observations}. We therefore will not consider the data set from the first 5 exposures {in the further analysis}. Besides, we checked that the discard of these data do not affect the main scientific results significantly.} The total number of stars with {\sl RVs} is {1880 from the spectra obtained through 2018 May 28-31}. Figure\,\ref{expo} shows the distribution of {the number of {\sl RV} determinations that was derived for each of these stars from these 25} exposures. We find that more than half of the targets have { 25} {\sl RV} measurements. The {\sl RV}s of the same stars visited multiple times can be an excellent practice to examine the robustness of {\sl RV}s {derived} from one system (or telescope). We calculate the {relative} {\sl RV}s ($\Delta${\sl RV}s) for each
targets by {subtracting the weighted mean of their values, where the square of ${S/N}$ was used as weight.}
Figure\,\ref{4W8} shows the scatter of the measured $\Delta${\sl RV}s. {From the} distribution we can be {directly} seen {that the precision is roughly 1\,km/s. However, the outlier measurements are possibly the results from {\sl RV} variables in particular at high $S/N$}.

\subsection{Selection of constant {\sl RV} stars}

To {precisely} check where the {outlier points} come from, {or concretely to estimate the precision,} we need to select the ``constant" {\sl RV} stars  first. Stars will { fall} into our sample if they have relative small $\Delta${\sl RV} from different plates.
{The concrete value is taken as 1\,km/s since it is the rough precision as estimated from Figure\,\ref{4W8}. In addition, we find that }more than half of 1{880} stars whose {\sl RV}s show standard deviation less than {1\,km/s}. This criterion can be more strict but the action will lose number of stars to compare the systematic errors in the following sections. The final sample contains {803} stars with {20075} {\sl RV}s, which are measured from all the {25} plates, called ``common constant" stars {below}.

\subsection{Analysis of systematic errors}
\begin{figure*}
\centering
\includegraphics[width=12.8cm]{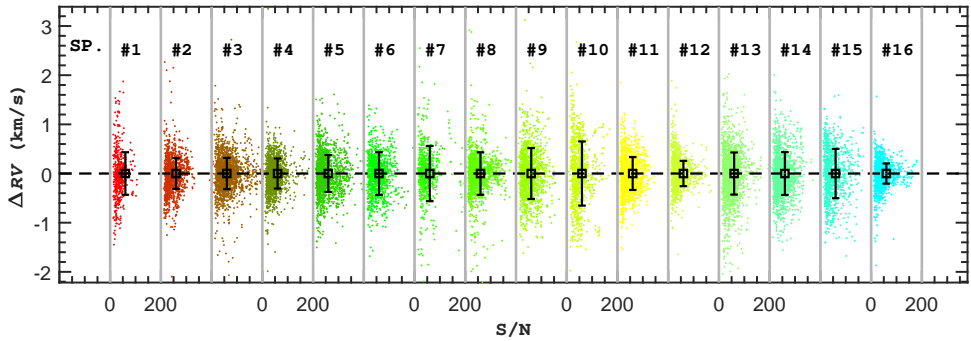}
\caption{{Distribution} of the relative {\sl RV}s ($\Delta RV$) {of} the ``common constant" stars as a function {$S/N$ (the IDs of spectrographs are}
marked in numbers on the upper of each panel).
{The $S/N$ scale between two consecutive vertical lines is set to be 200. } The horizontal dashed line represents the {\sl RV} under ideal measurement, that is zero, without any deviation. The weight values of each groups are given by open squares with their associated errors (standard deviations). More details are given in the text.
\label{rvsp}}
\end{figure*}

\begin{figure*}
\centering
\includegraphics[width=14cm]{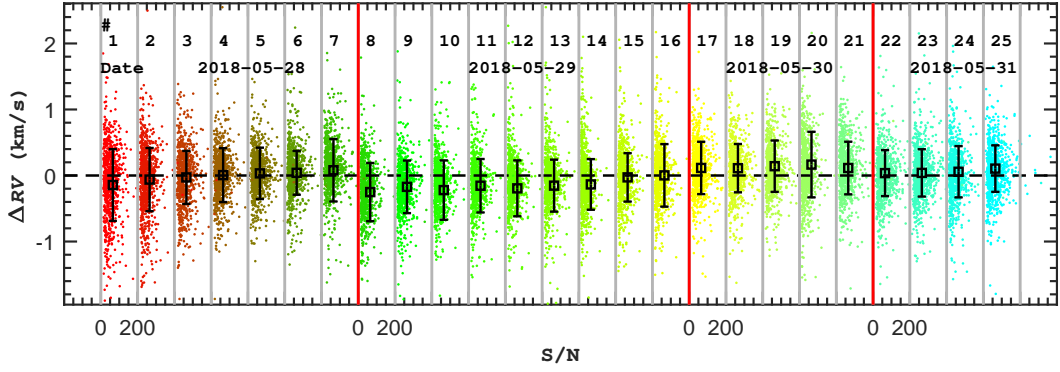}
\caption{Similar to Figure\,\ref{rvsp} but {according to the times of exposure}. The vertical {(red)
lines} indicate the exposure sequences in different {nights} marked by their dates {(UTC)} just below the {exposure} number.
\label{rvtime}}
\end{figure*}

Figure\,\ref{rvsp} shows the distribution of $\Delta RV$ where the common {constant} stars are divided into 16 groups as labeled by their spectrograph IDs. The results suggest {very small} systematic errors between different spectrographs,
as revealed by the {weighted values\footnote{The same weight is taken as the one mentioned in Section~\ref{s:drv}.}} of the $\Delta RV$. {The values are all near zero but with different standard deviation (see the errorbars in this figure). We note that the symbol itself shows a size of about 200\,m/s. The existence of systematic errors between different {plates is} illustrated in Figure\,\ref{rvtime}. The measured $RV$s {were now} divided into 25 groups labeled by the {sequence number of the observed plate}.
We clearly see that there are several $\Delta RV$ leaps between different nights (as indicated by red vertical lines), typically with {values} on the order of a few hundreds m/s. In addition, within the same night, {shifts are seen between consecutive plates} though {they are} smaller than {the} typical values between different nights generally}.


\subsection{Correction of systematic errors} \label{s:cse}
As shown in the {previous} section, systematic errors {exist} among the {\sl RV} measurements when they are obtained at different observational times {(major factor)} and from different spectrographs {(minor factor)}. These errors {induce an enlargement of} the {uncertainties} of {\sl RV} measurements from the LAMOST medium resolution spectra. In this section, we
introduce a technique to handle these systematic errors, which will significantly improve the
{\sl RV} precision.

We still use the common constant stars to correct the systematic errors. This time, all these
stars are divided into $25\times16$ groups by their plate ID and {spectrograph} ID. We calculate the averaged {weights}
$\overline{\Delta RV}_{ij}$ with the formula as
\begin{equation}
\overline{\Delta RV}_{ij} = \frac{\sum_k x_k \cdot \Delta RV_{ijk}}{\sum_k x_k},
\end{equation}
where the $x_k$ is the square of $S/N$, the index $k$ denotes the sequence of each {star} within one group which
is identified by its {indices} $i \in [1, 25]$ and $j \in [1, 16]$. $\overline{\Delta RV}_{ij}$ are the systematic errors since the { {\sl RVs} of the} common constant stars {are} independent {on} its observational time and
{spectrograph}. We can easily correct the systematic errors by {applying the formula}
\begin{equation}
RV_{corr} = RV - \overline{\Delta RV}_{ij},
\end{equation}
where the $RV$ {(with the omission of the subscript $i$, $j$, $k$)} is the measured radial velocity from the LAMOST pipelines.

\begin{figure}
\centering
\includegraphics[width=10cm]{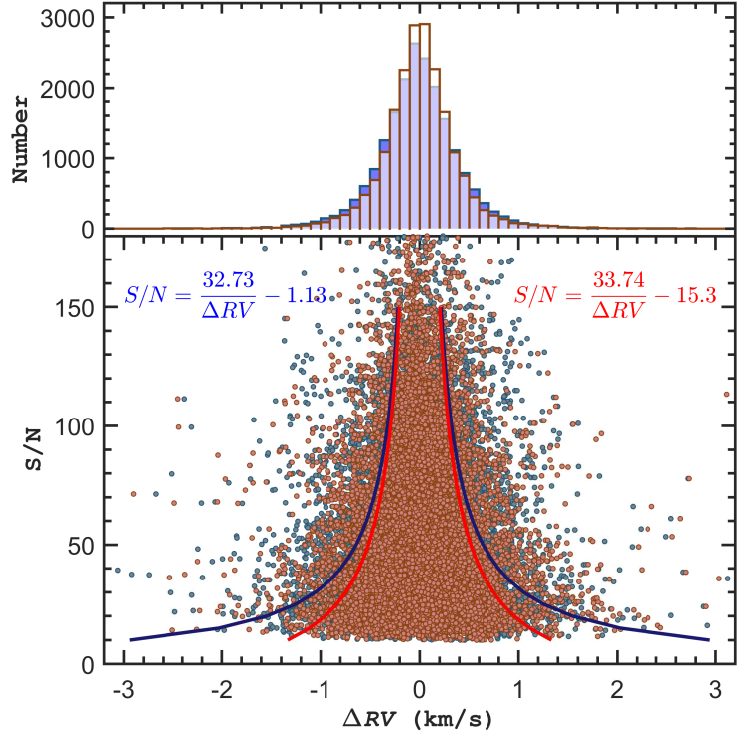}
\caption{Similar to Figure\,\ref{4W8} but for the constant stars before {(in blue)} and after {(in brown)} correction  of the systematic errors. The solid curves represent the optimal fitting whose function is given in the bottom panel (see text for details).
\label{rvs}}
\end{figure}

Figure\,\ref{rvs} shows the distribution of the $\Delta RV$s {before and} after {correcting} the systematic errors. The distribution of $\Delta RV$s now is unimodal centering around zero {with a slight shift of about 0.03\,km/s to its uncorrected values}, which suggests that the systematic errors have been corrected. The fitting curve shows that the precision of the {\sl RV} measurement is a function {of} the quality of {the} spectra ($S/N$). We note that the fitting {is} performed on the data with $S/N \in [10, 150]$ {since} the number of spectra with {a higher} $S/N$ {value} is very small and the outlier data points will greatly {affect} the fitting {of the} curve.
The 1$\sigma$ precision {reaches $\sim1.3$, $\sim1.0$, $\sim0.5$ and $\sim0.3$\,km/s} at $S/N=10$, 20, 50, and 100 {after the correction, instead of $\sim2.9$, $\sim 1.5$, $\sim0.6$ and $\sim0.3$\,km/s before the correction, respectively. This correction indicates that the precision will be especially improved for the spectra with $S/N < 50$.}

\section{Calibration of {\sl RV}s}
\subsection{External errors with APOGEE}
\begin{figure}
\centering
\includegraphics[width=8cm]{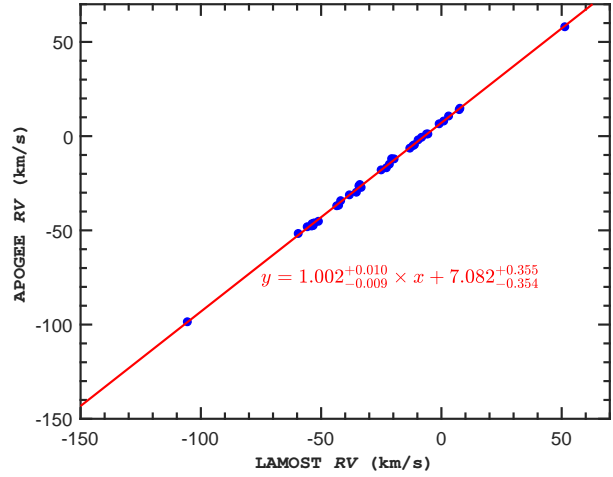}
\caption{Statistical comparison of radial velocity between LAMOST and APOGEE. {The best linear fit corresponds a line that is nearly parallel to the bisectrix.} Note that the error is smaller than the symbol itself.
\label{lap}}
\end{figure}

As we discussed the internal errors in the above section, in this section, we will discuss the comparison between LAMOST {\sl RV} {common constant stars} and APOGEE {\sl RV} standard stars.
We have cross-identified 34 stars with {\sl RV} measurements in our target list and from \citet{2018AJ....156...90H}, in a range from about {$-110$}\,km/s to $50$\,km/s. {We considered the {\sl RV} values after correction with equation\,(1)}. Figure\,\ref{lap} shows the statistical comparison for these 34 stars, where a good agreement between the two data sets can be clearly seen. The optimal fitting is {nearly} a parallel line to the {bisectrix} with a zero-point shift of about 7\,km/s.

\subsection{An scientific example of combination with photometry}
\begin{figure}
\centering
\includegraphics[width=9.2cm]{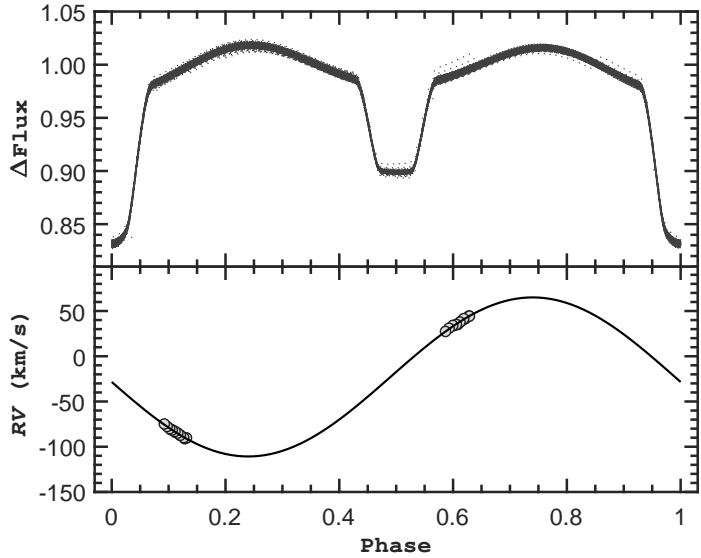}
\caption{The fold light curves (top panel) and {\sl RV}s (bottom panel)
of KIC~6863229 as a function of phase. The fitting curve in the bottom panel shows a sinusoidal wave (solid line).
\label{kic6863229}}
\end{figure}

After we determined the external and internal errors, the {\sl RV}s derived from medium-resolution spectra can be calibrated with enough precision. We here merely present one example of science cases where an eclipsing binary star with legacy data from {\sl Kepler} were observed by LAMOST. In this case, the mass of {the} binary components can be precisely determined \citep[see, e.g.,][]{2017ApJ...850..125Z}. KIC~6863229 is such a kind of stars, with $\alpha (2000) =19:31:02.82$, and $\delta(2000) +42:19:43.10$, and {\sl K}p = 12.134\footnote{http://archive.stsci.edu}. This star {has 25 {\sl RV} measurements from the} LAMOST medium-resolution spectra { provided here}. The light curves are collected from 2009 May 02 to 2013 May 11. Figure\,\ref{kic6863229} shows the two different phase {diagrams}. Both of the two curves are calculated with the following ephemeris formula
\begin{equation}
\mathrm{Min.}I = \mathrm{BJD}\,2454954.485(52)+ 1^\mathrm{d}.99492(28) \times E,
\label{epoch_KIC}
\end{equation}
where $T_0 = \mathrm{BJD}\,2454954.485(52)$ and $P = 1.99492$\footnote{There two values
can be found at http://keplerebs.villanova.edu/overview/?k=6863229.}\,(d)
are the time of a primary eclipse and {the available} period, respectively, {while} $E$ refers to the cycle number. A more detailed analysis of those data can be found in a forthcoming paper (Liu, et al. 2019).

\section{Discussion}
\label{sect:discussion}
The precision of {\sl RV}s from LAMOST medium-resolution spectra suffers {from slightly and significantly} systematic errors induced
by different spectrographs and {observation times, respectively}, particularly for the {observation} campaigns
{with} large gaps. The {most} significant systematic errors {are found between different observational nights, which
may have zero-point differences of about 0.5\,km/s}.
There also exists a slight drift for the {\sl RV} measurements during the same night, typically with a value of a few hundreds~m/s. The instrumental effects can account for that, such as the cooling
device which is put on the CCDs of LAMOST. {Due to fuel} consuming, the weight of that device will change and influence the position of spectra where their position is used for calibrating the wavelength. To avoid this, a semi-conductive devices will be used for cooling down the CCDs without changing their weight. The {slight} systematic errors between different spectrographs are very possibly caused
by zero-point differences between these spectrographs, thus, again, changing the wavelengths which {are} used for deriving {\sl RV}s.

Although the {\sl RV}s suffer {from} systematic errors, these errors can be corrected through different techniques. In this {paper}, we address one method to correct the measured {\sl RV}s and the results look reasonable. Our calculation based
on {803} common ``constant" stars which {have} {\sl RV}s not {changing} $>1$\,km/s over time. The systematic errors caused by instrument {effects} or observational campaigns should be {the} same to all {the} stars. Therefore, {one} can use these stars to evaluate the intrinsic precision of {\sl RV} measurement. Our results also give an estimation of {the} precision {for} different quality of medium-resolution spectra as indicated by their $S/N$ in {\sl SDSS}-like {\sl r} band. {After the correction, t}he precision {reaches $\sim1.3$, $\sim1.0$, $\sim0.5$ and $\sim0.3$\,km/s} at $S/N=10$, 20, 50, and 100, {which the corresponding values before correction are $\sim2.9$, $\sim 1.5$, $\sim0.6$ and $\sim0.3$\,km/s, respectively.}
Another technique is to calculate differential {\sl RV}s {before re-shifting} the {\sl RV}s' zero-points, which is very similar to {the measurement of}
differential magnitudes for variable stars in photometry (Pan et al. in Prep.). Our method should also draw one attention to the low-resolution spectra probably {suffer} from systematic
errors as well. However, time series plates are {only obtained for a low percentage of plates}. The
better way to remove systematic errors in low-resolution spectra can {be using} some standard {\sl RV} stars based on a similar technique.

The external errors of LAMOST are also calculated through {34} common stars with APOGEE catalog from \citet{2018AJ....156...90H}. We have found a systematic difference of {$\sim 7$\,km/s} between those two data sets. We {discussed} an example of an eclipsing binary star, whose  calibrated {\sl RV} curve with reasonable accuracy {was analyzed in combination} with the {\sl Kepler} photometry. This could be very useful to derive robust fundamental parameters for such stars, in particular for masses \citep{2017ApJ...850..125Z}.

\section{Conclusion}
A plate in the {\sl Kepler} field had been observed by LAMOST with the medium-resolution spectrographs and {produced through the most updated pipelines with {\sl RV}s}. These multiple visiting targets offer an opportunity to test the accuracy and precision of the {\sl RV}s derived from this new system. {By analyzing the 25 plates obtained through 2018 May 28 -31,} we find that there are systematic errors between different spectrographs and observational campaigns. However, these errors can be well removed by dividing the targets into different groups {according} to the two observational factors. The internal errors for {\sl RV}s {are found to be} with the values of { 1.3, 1.0, 0.5 and 0.3}\,km/s at $S/N=10$, 20, 50, and 100, respectively. We also compare our results with the APOGEE {\sl RV} standard stars and find the external error is about {7}\,km/s based on 34 common stars.

We end this {paper} with the remark that the precision of {\sl RV}s of medium-resolution spectra
is a fundamental measurement for the medium-resolution survey of LAMOST in the next five
years, as well as the atmospheric parameters. The scientific goals {that can be studied with such spectra} are built on these precision.

\normalem
\begin{acknowledgements}
The Guoshoujing Telescope (the Large Sky Area Multi-object Fiber Spectroscopic Telescope LAMOST) is a National Major Scientific Project built by the Chinese Academy of Sciences (CAS). Funding for the project has been provided by the National Development and Reform Commission. LAMOST is operated and managed by the National Astronomical Observatories, {CAS. We acknowledge the helpful discussion with Peter De Cat that improves the manuscript.} JNF acknowledges the support from the National Natural Science Foundation of China (NSFC) through the grant 11673003 and 11833002. WKZ {acknowledges the support from the China Postdoctoral Science Foundation through the grant 2018M641244 and} the LAMOST fellowship as a Youth Researcher which is supported by the Special Funding for Advanced Users, budgeted and administrated by the Center for Astronomical Mega-Science, Chinese Academy of Sciences (CAMS).
{XQC and YHH are supported by Key Research Program of Frontier Sciences, CAS, through the grant QYZDY-SSW-SLH007.}
\end{acknowledgements}

\end{document}